\documentstyle[12pt,aas2pp4]{article}

\begin{document}
 
\title{On the Consistency between the Outer Rotation Curve of
       the Galaxy and the Microlensing Rate 
       towards the Large Magellanic Cloud}

\author{Zong-Hong Zhu}
\affil{Institute of Theoretical Physics, 
        	Chinese Academy of Sciences, Beijing 100080, China}
\and
\author{Xiang-Ping Wu}
\affil{Beijing Astronomical Observatory,
		Chinese Academy of Sciences, Beijing 100080, China}

\begin{abstract}
	Recent determination of the outer rotation curve of the Galaxy
suggests that the total Galactic mass is likely to concentrate into a region 
of radius $\sim2R_0$. If this finding is correct, the previously estimated
optical depth of microlensing events towards 
the Large Magellanic Cloud (LMC) should be correspondingly reduced.
It appears that a truncated Galactic 
halo model instead of the standard isothermal one extending to LMC
accounts naturally for the reported microlensing rate 
by the MACHO collaboration. Such an explanation is indeed obvious 
yet has been overlooked in the past. 
\end{abstract}


More than dozen microlensing events towards the Large Magellanic Cloud (LMC)
have been so far detected by the MACHO 
and EROS collaborations (Alcock et al. 1997; Renault et al. 1997). 
An analysis of the first two year result of the MACHO collaboration   
gives rise to the  microlensing rate of $\tau=2.1^{+1.1}_{-0.7}\times10^{-7}$. 
This is only about half of the value $\tau_{G}=5\times10^{-7}$ for a 
standard spherical Galactic halo model (Paczy\'nski 1986; Griest 1991),
but is of an order of magnitude larger than the optical depth by 
the machos and stars in the LMC halo/disk itself 
(Wu 1994; Sahu 1994; Gould 1995).
Such a discrepancy, together with the puzzle of the too
large mass of $0.1$--$1M_{\odot}$ required for the machos in Galactic
halo, has led many authors to make attempt at modifying 
the conventional models for the Galactic halo and the macho distributions 
between the Milky Way and LMC (e.g. Ansari, et al. 1996; 
De Paolis et al. 1996; Zhao 1997; Gates et al. 1997).  It seems that many 
efforts are still needed before a better explanation of
the microlensing result and a clear understanding of  
structure of the Galactic halo are achieved.

Here, we propose an alternative yet natural way to reconcile the 
microlensing rates observed towards LMC with the measurement of outer 
rotation curve in the Galaxy. This is primarily motivated by 
the recent discovery of the Keplerian rotation curves of some galaxies
including the Milky Way (Honma \& Sofue 1996, 1997; Binney \& Dehnen 1997;
Ryder et al. 1997;
and references therein). These studies claimed that the declining rotation
curves are observed in a great number of galaxies, implying
that the possible truncation of galactic halos may occur at a certain radius
of several disk scale lengths. Detection of such Keplerian rotation 
curves challenges the traditional point of view about the mass distribution 
in galaxies, which may have a significant impact on the study of dark matter
in galactic halos including the ongoing searches for machos in the Galactic
halo. Recall that the flat rotation curves beyond the optical disks are
the unique convincing evidence for the presence of a large amount of
dark matter in galactic halos (see Ashman 1992 for a review). In particular,
the flat rotation curves indicate that the galactic halos can in principle
extend to rather a large distance ($>100$ kpc) with a density profile of 
$r^{-2}$. This uncomfortable ``standard'' model of dark galactic halo 
has been, nevertheless, accepted by astronomical society. Yet, it is a logical
and also natural result that the dark halos should be truncated 
at some radii $R_{cut}$, no matter how large $R_{cut}$ would be for various
galaxies.  In a sense, the recent detection of the Keplerian
rotation curves of galaxies are not beyond expectation. 

What would happen to the theoretical prediction of the optical depth
$\tau_{G}$ if the galactic halo shows a declining tendency at a radius
$R_{cut}$ ? The analysis by Honma \& Sofue (1996) 
and Binney \& Dehnen (1997) indicates that
$R_{cut}\approx2R_0$, where $R_0$ is the distance from the Sun to the
Galactic center. Their best-fit values of the 
the galactic constants $\Theta_0$, the circular velocity at $R_0$, and
$R_0$ are  $\Theta_0\approx200$ km s$^{-1}$ and $R_0\approx7.6$ kpc, 
respectively. 
These parameters are somewhat smaller than the IAU standard values. 
Applying these values for a truncated isothermal Galactic halo model 
up to $R_{cut}=15$ kpc instead of the traditional halo model extending
to the LMC, we find the optical depth of microlensing towards LMC to be  
only $\tau_{G} \approx 2\times10^{-7}$. This estimate should be regarded as  
a low bound on the optical depth since the rotation curve does not 
show an abrupt decline at $R_{cut}$. Rather, as argued by  
Honma \& Sofue (1996) and Binney \& Dehnen (1997),  
the Keplerian velocity decreases very
gently in the outer region, which may account for the fact that the flat
rotation curves often turn out to be a good approximation for most of
galaxies. In the case of the Milky Way, the circular velocity only reduces
to 190 km s$^{-1}$ even at $3R_0$. So, a robust estimate of the optical
depth of microlensing towards LMC depends on how a realistic model 
of the Galactic halo is constructed, which should take the variation
of the rotation velocity along radius into account. 
We will not explore the detailed model in the present letter.
Overall, a truncated Galactic halo model can easily provide an optical depth
that is consistent with the result by the MACHO collaboration.

Does a truncated Galactic halo model affect our estimate of the macho
mass ? To answer the question, we have computed the probability $P$ 
that a microlensing event of duration $T$ and maximum magnification
$\mu_{max}$ be produced by a macho with mass $m$ using the formulism
of Jetzer \& Mass\'o (1994). The maximum of $P$ gives rise to the 
most probably value of $mr_E^2/v^2T^2$, where 
$r_E^2=4GM_{\odot}D_s/c^2$, $D_s$ is the distance to the LMC and 
$v$ is the relative velocity of traverse motion of the macho. 
Numerical computation shows that the position of the maximum of $P$ 
shifts from $mr_E^2/v^2T^2=13.0$ to $17.9$ when the extent of the
Galactic halo is truncated at $2R_0$ instead of $D_s$. It turns out
that the estimate of the most probably value of macho mass is nearly
unaffected by the truncation of the Galactic halo if a circular
velocity of $\Theta_0=200$ km s$^{-1}$  (Honma \& Sofue 1996)
or $180$ km s$^{-1}$ (Olling \& Merrifield 1997) 
instead of $220$ km $s^{-1}$ 
is adopted. In other words, a truncated Galactic halo model does not
help to release the difficulty of the unreasonably large macho mass 
($\sim0.5M_{\odot}$) inferred from the current MACHO experiment. 

We emphasize in this brief note the consistency 
between the recent determination
of the outer rotation curve of the Galaxy (Honma \& Sofue 1996; 
Binney \& Dehnen 1997) and the observed microlensing rate towards
the LMC (Alcock et al. 1997). These measurements may significantly 
change our previous view of the dark Galactic halo. In any case,
the straightforward inference that machos contribute 
a fraction of $\sim50\%$ to the
halo dark matter based on the standard isothermal model
and the reported microlensing rate could be misleading.

\acknowledgements

This work was supported by the National Science Foundation of China.

%\clearpage
%\begin{thebibliography}{}

%\end{thebibliography}

\end{document}